\documentclass{sig-alternate-10pt}
\usepackage[margin=1in]{geometry}
\usepackage[T1]{fontenc }

\usepackage{xspace}
\usepackage{amsmath,amssymb}
\usepackage{xcolor}
\usepackage[hidelinks,breaklinks]{hyperref}
\usepackage{breakurl}
\usepackage{amsmath,tabu}
\usepackage{hhline}
\usepackage{cite}
\usepackage{cleveref}
\usepackage{balance}
\usepackage{subcaption}
\usepackage{graphbox}

\usepackage[utf8]{inputenc}
\usepackage{microtype}

\usepackage{mathptmx}
\usepackage{caption}
\usepackage{subcaption}
\usepackage{mathtools}
\usepackage{graphicx}
\usepackage{latexsym}
\usepackage{tabularx}
\usepackage{url}
\usepackage{booktabs}
\usepackage{paralist}

\usepackage{tikz}

\usepackage{titlesec}
\newtheorem{theorem}{Theorem}[section]

\newtheorem{example}[theorem]{Example}

\newtheorem{definition}[theorem]{Definition}

\NewDocumentCommand{\db}{}{\ensuremath{\mathbb{D}}}

\NewDocumentCommand{\tup}{O{x}}{\ensuremath{\mathbf{#1}}}

\NewDocumentCommand{\from}{}{\ensuremath{\colon}}
\RenewDocumentCommand{\to}{}{\ensuremath{\rightarrow}}

\NewDocumentCommand{\ecup}{}{\ensuremath{\mathbin{\bar{\cup}}}}
\NewDocumentCommand{\euplus}{}{\ensuremath{\mathbin{\bar{\uplus}}}}
\NewDocumentCommand{\dec}{}{\ensuremath{\mathsf{dec}}}

\NewDocumentCommand{\lit}{O{X}}{\text{Lit}(#1)}

\NewDocumentCommand{\sP}{}{\#P}
\NewDocumentCommand{\NP}{}{NP}
\NewDocumentCommand{\coNP}{}{coNP}

\NewDocumentCommand{\var}{m}{\ensuremath{\mathsf{var}(#1)}}
\NewDocumentCommand{\attr}{m}{\ensuremath{\mathsf{attr}(#1)}}
\NewDocumentCommand{\rel}{m}{\ensuremath{\mathsf{rel}(#1)}}
\NewDocumentCommand{\poly}{}{\ensuremath{\mathrm{poly\,}}}

\newcommand{\vtitle}{Tractable Circuits in Database Theory}
\title{\vtitle}

\numberofauthors{2}

\author{
\alignauthor Antoine Amarilli\\ %
 \affaddr{LTCI, Télécom Paris}\\
 \affaddr{Institut Polytechnique de Paris}\\
 \email{antoine.amarilli@telecom-paris.fr}
 \alignauthor Florent Capelli\\
 \affaddr{Université d'Artois, CNRS, UMR 8188}\\
 \affaddr{Centre de Recherche en Informatique de Lens}
 \email{capelli@cril.fr}
}

\hypersetup{
    colorlinks,
    citecolor={blue!50!black},
    urlcolor={blue!30!black},
    linkcolor={red!50!black},
    pdfauthor={Antoine Amarilli and Florent Capelli},
    pdftitle={\vtitle}
}

\begin{document}

\maketitle

\begin{abstract}
  This work reviews how database theory uses tractable circuit classes from
knowledge compilation. 
We present relevant query evaluation tasks, and notions of
tractable circuits.
We then show how these tractable circuits can be used to
address database tasks. We first focus on Boolean
provenance and its applications for aggregation tasks, in particular
probabilistic query evaluation. We study these for Monadic Second Order (MSO)
queries on trees, and for safe Conjunctive Queries (CQs) and
Union of Conjunctive Queries (UCQs). We also study circuit representations of query answers, and their
applications to enumeration tasks: both in the Boolean setting (for MSO) and the multivalued setting (for CQs and UCQs). 

\end{abstract}

\section{Introduction}
\label{sec:introduction}

The field of \emph{knowledge compilation}~\cite{darwiche2002knowledge}
studies how to efficiently reason about propositional knowledge bases,
and how to represent logic formulas as data structures that ensure the
tractability of tasks such as Boolean satisfiability.
These data structures are often based on \emph{decision
diagrams} such as OBDDs~\cite{bryant1986obdd} or on restricted versions of
Boolean \emph{circuits}; they often naturally correspond to the trace of an
algorithm~\cite{huang2005dpll}.
Such data structures can then 
be used to reason on the knowledge base, using different tools depending on
the task (satisfiability, model counting, etc.).

These tasks are naturally connected to database problems: e.g., satisfiability may be
seen as testing whether a query has an answer, and model counting can be seen as finding its number of answers. The main difference is that knowledge compilation focuses on propositional logic, which can be seen as the minimal setting where the techniques can be applied. It is hence only natural to use these techniques and tools from knowledge compilation and adapt them to the setting of databases.
In this paper, we review this line of work which designs efficient algorithms for
database tasks using tractable circuit representations of Boolean functions or
relations.

We identify two main ways in which circuit classes are used in database
theory. The first is via Boolean functions that naturally arise in databases,
such as \emph{Boolean provenance}, or \emph{answer functions} that represent the
output of queries. Knowledge compilation proposes efficient formalisms in which
we can represent such Boolean functions and ensure that some tasks
are tractable over them. Thus, if we can efficiently compute the Boolean
provenance or answer functions and represent them in a tractable circuit
class, then we can show some theoretical tractability results, and/or use
practical tools. This approach has been used across different domains, e.g.,
query evaluation on probabilistic databases~\cite{jha2012knowledge},
enumeration for monadic second-order queries~\cite{amarilli2017circuit}, and
Shapley value computation~\cite{deutch2022computing}; we will review these
works and more. Conversely, if we can show that the output to some problems
\emph{cannot} be tractably represented as a circuit in a given class, then we
show that these problems cannot efficiently be solved by algorithms whose trace
falls in that class: this approach is followed by~\cite{beame2017exact} among
others.

The second family of applications of circuit classes in databases is under the guise of
\emph{factorized databases}~\cite{olteanu2015size}, in which relations are
succinctly represented as circuits with restricted operators from
relational algebra.
This notion was introduced 
to understand
various tasks on the answer set of restricted kinds of Conjunctive Queries (CQs)~\cite{yannakakis1981algorithms,bagan2007acyclic,pichler2013tractable}. 
As we will explain, factorized databases can be seen
as natural generalizations of the circuits used in knowledge compilation, 
going from the Boolean domain to
a multivalued domain~\cite{olteanu2016factorized}.
We will then explain how such circuits can be used to recover known
results.

In both these settings, we see the strengths of
circuits: they give
a unifying view on scattered results, and they provide a modular and
generic way to solve complex tasks. Namely, they make it possible to design
query algorithms whose only task is to produce a circuit that falls in a
given class. The circuit can then be fed to existing algorithms and software implementations
which can solve various problems on the circuit independently from how it was built.

The paper is organized as follows.
\Cref{sec:bool-funct-datab} first reviews how \emph{Boolean functions}
naturally occur in database tasks. We focus on \emph{aggregation tasks} that
we relate to \emph{Boolean provenance}, and \emph{enumeration tasks} that we
relate to \emph{answer functions}. We then study in 
\Cref{sec:kc} how to efficiently represent such functions: we review known
circuit classes from knowledge compilation and the associated tractability
results and software implementations. We then explain how the tractable circuit classes of
\Cref{sec:kc}
allow us to
address the tasks of \Cref{sec:bool-funct-datab}: we start in \Cref{sec:mso} with 
Monadic Second-Order (MSO) queries on trees, then study aggregation
tasks for CQs and Union of CQs (UCQs) in \Cref{sec:dnnf-db}. 
We then move in \Cref{sec:query_answers} to the multivalued
perspective: we see how tractable circuit classes 
can be seen as factorized relations to
succinctly represent the 
answer set of CQs.
We close with an overview of other directions and questions for future
research
in \Cref{sec:perspective}.

%

%
%
%

%
%
%
%
%

%

%%% Local Variables:
%%% mode: latex
%%% TeX-master: "main"
%%% End:

\section{Database Tasks and Boolean Functions}
\label{sec:bool-funct-datab}

We first show in this section how various database tasks
can be expressed in terms of \emph{Boolean functions}, to be later
represented by circuits. We focus on two
main kinds of tasks. 
First, we present tasks expressed in terms of the \emph{Boolean
provenance} of a query on the input data.
We focus on
\emph{aggregation tasks} on the provenance, which intuitively
involve some form of counting.
Second, we will study how tasks can be expressed in terms of 
Boolean functions that
capture the answer of queries, e.g., in MSO.
We focus there on
\emph{enumeration tasks}, which ask for the
computation of witnesses.
We last sketch \emph{other tasks} related to circuits that we do not
investigate in detail.

\subsection{Boolean Provenance}

We first recall some standard terminology.
Provenance is defined over \emph{(relational) instances}, which 
are 
sets of facts over a signature. 
Formally, a \emph{signature} consists of a set of relation names with an
associated arity, a \emph{fact} for a relation name $R$ of arity~$n$ is an
expression of the form $R(a_1, \ldots, a_n)$ with $a_1, \ldots, a_n$ being
values, and an \emph{instance} is a set of facts.
A \emph{Boolean query} $Q$ is then a function
that maps instances~$\db$ to a Boolean value indicating whether the query is
satisfied by~$\db$. The \emph{Boolean provenance} of~$Q$ over an
instance~$\db$ then simply describes the truth status of~$Q$ on all
\emph{subinstances} of~$\db$, i.e., all subsets of its facts. Formally:

\begin{definition}
  \label{def:prov}
  Let $\db$ be an instance and $Q$ be a Boolean query. The
  \emph{provenance of~$Q$ on~$\db$} is the Boolean function from~$2^\db$ to~$\{0,
  1\}$ that maps each subinstance $\db' \in 2^\db$ of~$\db$ %
  to~$1$ if~$Q$ is true on~$\db'$ and to~$0$ otherwise.
\end{definition}

As 
a Boolean function, provenance can be represented in multiple
ways, e.g.,
as \emph{Boolean formulas} like in the example below, or
as \emph{Boolean circuits} (see Section~\ref{sec:kc}).

\begin{example}
  \label{exa:prov}
  Let $\db = \allowbreak \{R(a), R(a'), S(b)\}$ be an instance
  and $Q$ be the Boolean CQ $\exists x \, y ~ R(x), S(y)$ asking for the
  presence of an~$R$-fact and of an~$S$-fact.
  The provenance of $Q$ on~$\db$ is the
  function denoted by the Boolean formula $(R(a) \lor R(a')) \land S(b)$.
\end{example}

\paragraph*{Semiring provenance and more general semirings.}
Boolean provenance as defined here is the special case of
\emph{semiring provenance}~\cite{green2007provenance} for the
semiring $\text{Bool}[X]$ of Boolean functions.
One advantage of Boolean provenance is that
it is purely
semantic, i.e., it considers the query as a black-box. By contrast, provenance
for more general semirings like $\mathbb{N}[X]$ often depends on how the
query is executed.
We note that circuit notions have also been defined for such general provenance
semirings~\cite{deutch2014circuits,amarilli2015provenance}, in connection with
arithmetic circuits;
but we only discuss Boolean provenance from now on.

Now that we have defined Boolean provenance for queries, we explain how database
tasks can be rephrased in terms of Boolean provenance. We will focus here on
\emph{aggregation tasks}, where we intuitively want to
perform a kind of counting over subsets of the input instance.

\paragraph*{Uniform reliability.}
The simplest aggregation task is \emph{uniform reliability} (UR) for a
Boolean query~$Q$:
given an instance $\db$, count the subinstances of~$\db$ that satisfy~$Q$.
This can be solved via \emph{model counting}. Formally:

\begin{definition}
  Let $\phi$ be a Boolean function over variables~$X$. The \emph{model counting}
  problem (\#SAT) for~$\phi$ asks how many valuations of~$X$ satisfy~$\phi$.
  Formally, for $Y \in 2^X$, we write $\nu_Y$ for the Boolean valuation over~$X$
  that maps $x \in X$ to~$1$ if~$x \in Y$ and to~$0$ otherwise.   \#SAT
  is the problem of computing
  $\#\phi := |\{Y \in 2^X \mid \nu_Y \text{~satisfies~} \phi\}|$.
\end{definition}

Hence, if $\phi$ if the provenance of~$Q$ on an instance~$\db$,  then
  the answer to UR for~$Q$ is the
  answer to \#SAT for~$\phi$.

\paragraph*{Probabilistic query evaluation.}
A generalization of UR is \emph{probabilistic query evaluation} (PQE) for~$Q$. In this
setting, we are given a so-called \emph{tuple-independent database} (TID): it
consists of an instance~$\db$ with a function $\pi\colon\db\to [0, 1]$
mapping each fact of~$\db$ to a rational probability value. We assume
independence across facts, and consider the product probability distribution on
subinstances of~$\db$, where the probability of~$\db' \subseteq \db$ is
$\pi(\db') := \prod_{F
\in \db'} \pi(F) \times \prod_{F \in \db \setminus \db'} (1 - \pi(F))$. We want to
know the total probability of the subinstances of~$\db$ that satisfy~$Q$,
namely, $\sum_{\db'\subseteq\db, \db' \text{~satisfies~} Q}
\pi(\db')$. This problem can be solved via \emph{weighted model counting}:

\begin{definition}
  \label{def:wmc}
  Let $(K,\otimes,+)$ be a semiring, and
  let $\phi$ be a Boolean function over~$X$. We write $\lit$ the set of
  literals over~$X$, i.e., $X \cup \{\neg x \mid x \in X\}$.
  Let $w \colon \lit \to K$ be
  a \emph{weight function} giving a weight in~$K$ to each literal over~$X$.
  The
  \emph{weight} of a Boolean valuation $\nu\colon X \to \{0, 1\}$ is then
  $w(\nu) := \bigotimes_{x \in X, \nu(x) = 1} w(x) \times \bigotimes_{x\in X, \nu(x) = 0} w(\neg
  x)$. Then, the
  \emph{weighted model counting} problem (WMC) for~$\phi$ is to compute the total
  weight of 
  satisfying valuations, i.e., $\sum_{\nu \text{~satisfies~} \phi}
  w(\nu)$.
\end{definition}

Note that
model counting for~$\phi$ reduces to WMC with a weight of~1 for each literal.
We then have that PQE for~$Q$ on a TID $(\db, \pi)$ amounts to WMC for the provenance~$\phi$
of~$Q$ on~$\db$ in the semiring of rationals $(\mathbb{Q},\times,+)$ with weights given by~$\pi$.

As the tasks UR and PQE are often intractable, we may prefer to
study \emph{approximate model counting} (ApproxMC) and its weighted variant
(ApproxWMC). In these variants, instead of solving the
problem exactly, we wish to compute an approximation of the model count or of
the probability. We focus on \emph{multiplicative approximations}:
given an error $\epsilon > 0$, we must compute an approximation $\bar{x}$ of the true
value $x$ that ensures $(1-\epsilon) x \leq \bar{x} \leq (1+\epsilon) x$. We often allow approximation algorithms to be \emph{randomized}:
then, the output of the algorithm must be a correct approximation with 
probability at least $2/3$. We say that we have a \emph{fully polynomial-time
approximation scheme} (FPRAS) for Approx(W)MC when we have a
randomized algorithm to compute a multiplicative approximation of the count with
running time polynomial in the input instance and in the inverse of the desired
error~$\epsilon$.

\paragraph*{Shapley values.}
Another aggregation task 
is the computation of \emph{Shapley values}, which can be used to quantify the
contribution of a fact to making the query true~\cite{bertossi2023shapley}. In
this setting, we fix a query~$Q$ and we are given as input an
instance~$\db$, which is partitioned between so-called
\emph{exogenous} facts~$\db_{\mathrm{x}}$, which are always present, and
\emph{endogenous}
facts~$\db_{\mathrm{n}}$, which may be present or absent.
The Shapley value is then an aggregate over all subinstances of~$\db$ that
contain all the exogenous facts of~$\db_{\mathrm{x}}$. We
omit the formal definition of the Shapley value; see~\cite{bertossi2023shapley}.
Note that the Shapley value reduces in particular to counting
how many subsets of endogenous facts of a given cardinality satisfy~$Q$
together with the exogenous facts~\cite{livshits2021shapley}.

The computation of Shapley values can be posed on the Boolean provenance $\phi$ of the query~$Q$,
but imposing that all exogenous facts are kept.
This amounts to partial evaluation of~$\phi$, namely, setting the variables
of $\db_{\mathrm{x}}$ to~$1$.

\subsection{Boolean Answer Functions}

We now move on from Boolean provenance to a different kind of Boolean functions,
this time defined to represent the results of queries.
It will be easier to define these functions for queries with one free
second-order variable, i.e., queries that return subsets of domain elements as
answers, in particular queries expressed in MSO.
We will revisit this perspective in Section~\ref{sec:query_answers}
to work on more conventional query results that consist of relations.

Formally, the semantics of a query $Q(X)$ with a free second-order variable is that $X$ stands for a set of
elements taken from the active domain: given an instance~$\db$, letting $D$ be the
\emph{domain} of~$\db$ (the set of elements occurring in facts of~$\db$), the
\emph{answers} of~$Q$ on~$\db$ is the set of subsets $A \subseteq D$ such that
$Q(X := A)$ is true.

\begin{example}
  \label{exa:indep}
  On a signature with one binary relation $R$ and one unary relation~$V$, consider
  the query $Q(X) := \neg (\exists y\, z ~ (R(y, z) \lor R(z, y)) \land V(y)
  \land V(z) \land X(y)
  \land X(z))$. Then,
  given an instance $\db$ with domain~$D$, we have that $Q(X := A)$ holds
  precisely when $A \subseteq D$ is an independent set of the 
  graph
  represented by~$\db$ with vertices coded by~$V$ and edges coded by~$R$.
\end{example}

In this setting, for a query~$Q$, given an instance $\db$ with domain~$D$, we can
naturally define a Boolean function $\phi$ describing the answers of~$Q$
on~$\db$, called the \emph{answer function} of~$\phi$ on~$\db$. Formally, the
variables of the answer function
are the domain elements in~$D$, and
it is satisfied by a Boolean valuation $\nu\colon D\to \{0, 1\}$ precisely
when the set $A_\nu := \{a \in D \mid \nu(a) = 1\}$ is an answer
to~$Q$.

For a query $Q(X_1, \ldots, X_k)$ with multiple
second-order variables, we can also define the answer
function of~$Q$ on an instance $\db$ with domain~$D$: it is a Boolean function
with
variable set 
$D \times [k]$, satisfied by
the valuations $\nu\colon D \times [k] \to \{0, 1\}$ such that the tuple of the
sets $A_i := \{a \in D \mid \nu((a, i)) = 1\}$ for $i \in [k]$ is an answer
to~$Q$.

We now study how problems over the set of answers of a query~$Q$ can be posed over
their Boolean answer function. One example is \emph{answer counting}: the number of
answers of~$Q$ over an instance~$\db$ is the number of satisfying
assignments of the answer function of~$Q$ on~$\db$. However, in this section, we focus on
\emph{enumeration tasks}: given $\db$,
we must compute a set of solutions of~$Q$ over~$\db$.

\paragraph*{Finding and enumerating satisfying valuations.}
The simplest task on an answer function~$\phi$ is simply to decide if the query has an
answer, i.e., if $\phi$ has a satisfying valuation
(called \emph{satisfiability} or SAT); and to compute one if it exists.
A variant 
is \emph{uniform sampling}, i.e., a randomized algorithm
that must produce one satisfying valuation uniformly at random.
We note that the task can be made approximate
by allowing a \emph{failure probability} or doing
\emph{near-uniform sampling}.

However, it may also
be important to \emph{compute all satisfying valuations}.
One challenge to formalize the task is that the number of answers to produce can be large, which makes it
difficult to measure efficiency.

\begin{example}
  Continuing Example~\ref{exa:indep},
  consider the task of enumerating all query answers, i.e., all independent sets
  of the input graph.
  A naive algorithm to
  compute the answers is to list each possible subset~$A$ and test if it is an
  independent set. The
  naive algorithm 
  takes time $\Omega(2^n)$, but 
  it is hard to improve on this, because the
  worst-case complexity of the
  task is also $\Omega(2^n)$. Indeed,  the running time cannot be
  less than the number of answers to produce, and
  given an instance $\db$ with $n$ facts, there may be up
  to~$2^n$ answers (e.g., for 
  $n$ isolated vertices).
  What we intuitively want is 
  to beat
  $\Omega(2^n)$ on the instances where the answer set is small; or to produce
  the first few solutions faster than $\Omega(2^n)$.
\end{example}

This example illustrates why 
we study algorithms in terms
of the output size. One way 
is 
\emph{enumeration algorithms}, which have been studied in many contexts (see,
e.g., \cite{wasa2016enumeration}).
In data management, enumeration algorithms 
distinguish two phases: a \emph{preprocessing phase} to
perform some precomputations, and an \emph{enumeration phase} to produce
all solutions with small \emph{delay} between consecutive solutions. We omit the
formal definition of
enumeration algorithms; see, e.g.,
\cite{segoufin2014glimpse}.

\begin{definition}
  Given a Boolean function~$\phi$ on variables~$X$, the task of \emph{enumerating satisfying
  valuations} (Enum) of~$\phi$ asks us to efficiently produce the list of all
  satisfying valuations, with small preprocessing time and small delay between
  any two successive valuations. Note that, in the enumeration, we may write each
  valuation $\nu\colon X \to \{0, 1\}$ as the set $\{x \in X \mid \nu(x) =
  1\}$: this
  may be more concise 
  when the Hamming weight of~$\nu$ is small.
\end{definition}

Thus, for the database tasks presented above, we can devise enumeration
algorithms for queries via answer functions: compute a representation of the
answer function, then enumerate its satisfying valuations.

\paragraph*{Ranked enumeration and ranked access.}
Beyond the mere listing of satisfying valuations, it is often relevant to list
them \emph{in a specific order}, or to find the \emph{top-$k$ satisfying
valuations} in this order.
We also want to do \emph{ranked access}:
quickly access the $i$-th solution in this order, which generalizes
tasks such as quantile computation~\cite{carmeli2023accessing}.

\begin{definition}
  Given a Boolean function~$\phi$ on variables~$X$, given a total order~$<$
  defined on the set $2^X$ of valuations of~$X$, the task of
  \emph{ranked enumeration of satisfying valuations} of~$\phi$
  under~$<$ is to enumerate the satisfying valuations of~$\phi$ in the
  order given by~$<$. The task of computing the \emph{top-$k$ satisfying
  valuations}, given $k \in \mathbb{N}$, is to compute the $k$ satisfying
  valuations that are first according to~$<$. The task of \emph{ranked access}
  is to return, given $i \in \mathbb{N}$, the $i$-th satisfying valuation in
  the order,
  or fail if the number of %
  satisfying valuations is less than $i$.
\end{definition}

There are two main ways in which we can represent an order on valuations without
materializing its comparability graph explicitly. One first method is 
\emph{lexicographic orders}: we are given a total order ${<'}$ on the set~$X$ of
variables, and extend it lexicographically to valuations. Formally, given two
valuations $\nu, \nu' \in 2^X$, we have $\nu < \nu'$ if, letting $x$ be the
smallest variable for~${<'}$ such that $\nu(x) \neq \nu'(x)$, we
have $\nu(x) = 0$ and $\nu'(x) = 1$. The second method is 
\emph{weights}:
we fix a semigroup $(K,\otimes)$, and define the weight
$w(\nu)$ of a valuation~$\nu$ 
for a weight function $w$
like in Definition~\ref{def:wmc}.
Now, if we pick
some total order ${<'}$ on semigroup values, we can
sort valuations according to their weight in~$K$: we may typically assume that
${<'}$ is compatible in some sense with the semigroup law (e.g.,
\emph{subset-monotonicity}~\cite{amarilli2024ranked,tziavelis2022any}), and often impose some
tie-breaking rule to make the order total. In particular, when $K=\mathbb{Q}$
and weights are
probabilities (like for PQE), we can compute the \emph{most probable
satisfying valuation} or the \emph{most probable falsifying valuation}.

\subsection{Other Tasks via Boolean Functions}

Boolean functions can be used for many data management tasks beyond the ones
presented so far. We close the section by briefly alluding to other
such tasks.

First, the notion of \emph{database
repairs}, specifically in the case of \emph{maximal subset repairs},
 can be studied via Boolean provenance by looking at the maximal satisfying
 valuations of the provenance.
Thus, the  problem of
\emph{subset repair counting}~\cite{kimelfeld2020counting,calautti2019counting,calautti2022counting} can be seen as the
aggregation task of counting the maximal satisfying valuations of the provenance
of a certain Boolean query; and the task of
\emph{enumerating subset repairs}~\cite{kimelfeld2020counting} amounts to the
enumeration task of listing the maximal satisfying valuations.
A connection could also be phrased for \emph{consistent query
answering}~\cite{bertossi2006consistent,bertossi2019database}.

Second, more generally, the setting of \emph{reverse data management}~\cite{meliou2011reverse}
provides other examples of problems that can be phrased in terms of
Boolean provenance.
These problems can consider ways to delete facts in an input instance in order
to satisfy a property, for instance \emph{minimum
witness}~\cite{miao2019explaining,hu2024finding} (delete as many facts
as possible while making the query true) or
\emph{resilience}~\cite{freire2015complexity,makhija2023unified} (delete as few
facts as possible to make the query false). These problems can be posed on the
Boolean provenance: 
find satisfying valuations of small Hamming weight,
or falsifying valuations of large Hamming weight.

Third, in the setting of \emph{uncertain
databases}~\cite{abiteboul1995foundations}, we may consider
instances where some facts are \emph{uncertain}, i.e., they may be
present or absent; and other facts are \emph{certain} and are necessarily present.
These correspond to endogenous and exogenous facts in the case of Shapley value
computation. We can then study tasks such as \emph{certain query
answers}~\cite{imielinski1984incomplete}, i.e., is the query always true on all
possible worlds of an uncertain instance, i.e., is the Boolean provenance
tautological.
A related task is \emph{query-guided
uncertainty resolution}~\cite{drien2023query}
which interactively determines the truth value of the query, by 
probing individual facts while optimizing, e.g., the worst-case decision tree
height.
This amounts to 
\emph{stochastic Boolean function
evaluation}~\cite{allen2017evaluation} of the Boolean provenance.

Fourth, many of the tasks defined so far can be extended to \emph{incremental
maintenance}: compute the result (or its representation, e.g., as an
enumeration index), and maintain this result
efficiently under \emph{updates} to the data.
In general, circuit representations
are amenable to incremental maintenance:
we can reevaluate the circuit on a different
valuation whenever the data changes.
However, this idea
assumes that the initial set of facts never grows,
i.e., facts can only be removed or added back.

\section{Knowledge Compilation}
\label{sec:kc}

We have seen in \Cref{sec:bool-funct-datab} that many database tasks can be
rephrased to problems on Boolean functions.
However, these reductions do not 
directly give 
tractable algorithms, because many
of these problems are hard.
For example, SAT is \NP-complete already for Conjunctive Normal
Form formulas (CNFs)~$\phi$.
Computing $\#\phi$ is even harder:
it is
\sP-complete~\cite{valiant1979complexity},
and \NP-hard to
approximate to a $2^{n-\varepsilon}$-factor for every $\varepsilon>0$, even %
if $\phi$
has no negation and clauses of size at most $2$~\cite{roth1996hardness}. 
However, tractability can hold for other representations of Boolean
functions.
For instance, for a Disjunctive Normal Form (DNF) formula $\phi$,
finding a satisfying valuation is easy.
While
computing $\#\phi$ is also \sP-complete for DNFs, 
we can approximate it to a $(1\pm\varepsilon)$ factor with high probability
via a Monte-Carlo-based FPRAS, namely, the Karp-Luby algorithm~\cite{karp1989monte}.

Thus, one approach to make the tasks of \Cref{sec:bool-funct-datab} tractable
is to build
representations of the Boolean functions that are tractable for the task at hand.
This modular approach often makes it possible to show several tractability
results at once, and 
allows us to use efficient practical 
implementations 
(see
\Cref{sec:tools}).

To achieve this, we turn to the field of \emph{knowledge
compilation}~\cite{darwiche2002knowledge}, which
studies 
tractable representations for Boolean functions and their properties.
We review definitions, results, and tools from this field in this section,
which we will use 
in \Cref{sec:mso,sec:dnnf-db,sec:query_answers}
for various database tasks featuring aggregation and enumeration.

\subsection{DNNF Circuits}
\label{sec:dnnf}

Most representations of Boolean functions studied in the literature are
restricted forms of \emph{Boolean circuits}. The use of circuits allows for
efficient sharing and factorization, while the restrictions ensure
tractability. One of the most general such 
representation 
is Decomposable Negation Normal Form (DNNF) circuits~\cite{darwiche2001decomposable}.  We start by defining Boolean circuits before introducing DNNFs.

\paragraph*{Circuits and NNF circuits.}
Formally, a \emph{Boolean circuit} $C$ on variables $X$ is a Directed Acyclic Graph together with a distinguished node called the \emph{output}.
The nodes with indegree $0$, called the \emph{inputs of the circuit}, are each labeled either with a constant $1$ or $0$ or with a variable $x \in X$.
The internal nodes of~$C$, called \emph{gates}, are labeled by $\wedge$, $\vee$ or $\neg$. Given a gate $v$, every gate $w$ 
that has an edge to $v$ is called an \emph{input of $v$}. Every $\neg$-gate has exactly one input. 
Given a gate $v$ of $C$, we let $\var{v}$ be the set of variables $x \in X$ that appear in the subcircuit rooted in $C$, that is, we can reach~$v$ by a directed path
from an input of~$C$ labeled by~$x$.
The size $|C|$ of a Boolean circuit is the number of edges in the %
DAG.

Every gate $g$ of $C$ computes a Boolean function $f_g$ on $\var{g}$: given a valuation $\nu\colon\var{g}\to\{0, 1\}$, the value of~$f_g$ is defined by setting the inputs of the circuit with~$\nu$ and evaluating 
internal gates inductively up to $g$.
The Boolean function on~$X$ computed by~$C$ is then the function $f_o$ on~$\var{o}$ for~$o$ the output of~$C$, extended to~$X$ by allowing the variables of $X \setminus \var{o}$ to take any value.

\begin{figure}
  \centering
  \begin{subfigure}{.25\textwidth}
    \centering
    \includegraphics[align=c,width=4cm]{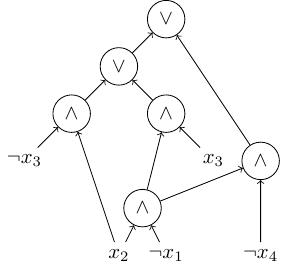}
  \end{subfigure}
  \begin{subfigure}{.2\textwidth}
    \centering
    {
      \renewcommand{\arraystretch}{0.88}
  \begin{tabu}{llll}
    \toprule    \rowfont[c]{\bf } $x_1$ & $x_2$ & $x_3$ & $x_4$ \tabularnewline \midrule
    0 & 1 & 0 & 0  \tabularnewline
    0 & 1 & 0 & 1  \tabularnewline
    0 & 1 & 1 & 0  \tabularnewline
    0 & 1 & 1 & 1  \tabularnewline
    1 & 1 & 0 & 0  \tabularnewline
    1 & 1 & 0 & 1  \tabularnewline

    \bottomrule
  \end{tabu}
}
\end{subfigure}

\caption{A DNNF and its satisfying valuations}
\label{fig:dnnf}
\end{figure}

We focus on Boolean circuits in \emph{Negation Normal Form} (NNF) where the input of every negation gate is an input of the circuit; 
alternatively, NNF circuits are circuits built on literals using $\wedge$ and $\vee$ on internal gates.
See \Cref{fig:dnnf} for an example of an NNF circuit.
Any Boolean circuit can be 
put in NNF in linear time 
using De Morgan's laws iteratively
to push negations down.

\paragraph*{DNNF circuits.}
We now move to restrictions on NNF circuits aimed at enforcing tractability
of problems (e.g., SAT).
Imposing NNF is not sufficient for this: as we explained, it is essentially without loss of generality.
Intuitively,
what makes SAT hard is that finding 
satisfying valuations $\nu_1$ for $f_1$ and $\nu_2$ for $f_2$ does not help to 
find one for $f_1 \wedge f_2$, because $\nu_1$ and $\nu_2$ may be inconsistent
on their shared variables. To avoid this, we use \emph{decomposable}
$\wedge$-gates.
A $\wedge$-gate $g$ is 
\emph{decomposable} if for every pair of distinct inputs $g_1,g_2$ of $g$, we
have $\var{g_1} \cap \var{g_2} = \emptyset$. A \emph{Decomposable Negation
Normal Form (DNNF)} circuit is an NNF circuit~$C$ where every $\wedge$-gate is decomposable. 
Observe that decomposability makes SAT tractable: we can propagate
satisfying valuations upwards in~$C$, building satisfying valuations for
decomposable $\wedge$-gates by concatenating satisfying valuations of their
inputs.
The circuit depicted on \Cref{fig:dnnf} is a DNNF. 

Note that, in database terms, a decomposable $\wedge$-gate~$g$ is a Cartesian product:
if we see the Boolean function computed by each input $g_i$ of~$g$ 
as a relation $R_i$ over attributes $\var{g_i}$ and domain $\{0,1\}$, then
the relation computed by $g$ is $R_1 \times R_2$. We revisit this in
\Cref{sec:query_answers}.

DNNF circuits clearly generalize DNF formulas, while retaining 
some of their tractability. Indeed, 
we can solve SAT on a DNNF circuit~$C$ in time $O(|C|)$, and
enumerate the satisfying assignments with a delay of $O(n|C|)$ where $n = |X|$. %
DNNFs are also closed under partial evaluation,
called \emph{conditioning} in knowledge compilation~\cite{darwiche2002knowledge}:
given a DNNF circuit $C$ on variables $X$ computing $f$ and a partial valuation $\nu$ on variables $Y \subseteq X$, we can build in time $O(|C|)$
a DNNF
computing $f|_{\nu}$.
Indeed, 
simply replace each node labeled with a variable~$x \in Y$ by the
constant~$\nu(x)$:
this does not affect decomposability.

\subsection{Restrictions on DNNF Circuits}

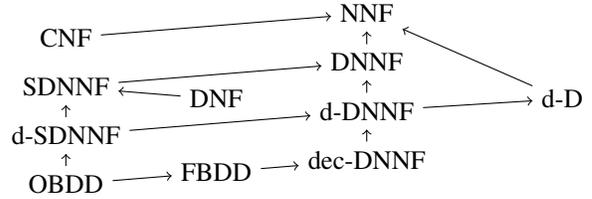
\begin{figure}
  \begin{tikzpicture}[xscale=4,yscale=.65]
    \node (obdd) at (0, 1) {OBDD};
    \node (fbdd) at (0.5, 1.25) {FBDD};
    \node (decdnnf) at (1, 1.5) {dec-DNNF};
    \node (dsdnnf) at (0, 2) {d-SDNNF};
    \node (sdnnf) at (0, 3) {SDNNF};
    \node (ddnnf) at (1, 2.5) {d-DNNF};
    \node (dnnf) at (1, 3.5) {DNNF};
    \node (dnf) at (.5, 2.75) {DNF};
    \node (dd) at (1.65, 2.75) {d-D};
    \node (nnf) at (1, 4.5) {NNF};
    \node (cnf) at (0, 4) {CNF};
    \draw[->] (obdd) -- (fbdd);
    \draw[->] (fbdd) -- (decdnnf);
    \draw[->] (decdnnf) -- (ddnnf);
    \draw[->] (obdd) -- (dsdnnf);
    \draw[->] (dsdnnf) -- (ddnnf);
    \draw[->] (dsdnnf) -- (sdnnf);
    \draw[->] (sdnnf) -- (dnnf);
    \draw[->] (ddnnf) -- (dnnf);
    \draw[->] (cnf) -- (nnf);
    \draw[->] (dnnf) -- (nnf);
    \draw[->] (dnf) -- (sdnnf);
    \draw[->] (ddnnf) -- (dd);
    \draw[->] (dd) -- (nnf);
  \end{tikzpicture}
  \caption{Inclusions between circuit classes. Arrows denote inclusion (i.e.,
  linear-time transformations); they are all known
  to be
  strict in the sense that reverse arrows do not exist,
  except inclusions involving \mbox{d-Ds} which are not separated from d-DNNF~\cite{monet2020solving}
  and only conditionally separated from NNF.
  Most results are in~\cite{amarilli2020connecting,amarilli2024circus}.%
  }
\label{fig:dnnf-hierarchy}
\end{figure}

We now introduce
additional restrictions to make more tasks tractable.
\Cref{fig:dnnf-hierarchy} and \Cref{fig:kc-task} summarize the circuit
classes studied and the complexity of tasks.

\paragraph{Determinism.}
Unlike enumeration, aggregation tasks such as \#SAT are not tractable on DNNFs, as they are already hard on DNFs.
Intuitively, \#SAT is hard because $\#f_1$ and $\#f_2$ does not give $\#(f_1
\vee f_2)$: indeed, $f_1$ and $f_2$ may be sharing some of their satisfying valuations.
\emph{Determinism}~\cite{darwiche2001tractable} is a way to make such tasks tractable.
In a Boolean circuit, a $\vee$-gate $g$ is said to be \emph{deterministic} if for every pair of distinct inputs $g_1,g_2$ of $g$, the function $f_{g_1} \wedge f_{g_2}$ is not satisfiable. In other words, if $\nu$ is a satisfying valuation of $f_g$, then there is exactly one input $g_i$ of $g$ such that $\nu|_{\var{g_i}}$ is a satisfying valuation of $g_i$. A \emph{deterministic Decomposable Negation Normal Form} (d-DNNF) circuit is a DNNF 
where every $\vee$-gate is deterministic.
The DNNF in \Cref{fig:dnnf} is \emph{not} a d-DNNF.

Determinism makes it possible to count satisfying valuations bottom-up in time
$O(|C|)$, assuming that arithmetic operations take constant time.
This counting task is easier to perform when assuming that the circuit is \emph{smooth}, i.e.,
that we have $\var{g'}=\var{g}$ for every input $g'$ of a $\vee$-gate $g$.
Smoothing can be enforced up to multiplying the circuit size by a
$|\var{C}|$ factor,
which can sometimes be improved~\cite{shih2019smoothing}. %
When a circuit $C$ is deterministic and smooth, we can compute bottom-up the number of
satisfying valuations of each gate: we have $\#(f_1 \land f_2) = \#f_1
\times \#f_2$ thanks to decomposability, and $\#(f_1 \lor f_2) = \#f_1 +
\#f_2$ thanks to determinism and smoothness.
Determinism and smoothness also ensure the tractability of WMC for weights in any semiring $(K,\oplus,\otimes)$~\cite{kimmig2017algebraic}, using only 
$O(|C|)$ semiring operations.
% , provided that the circuit is \emph{smooth}, i.e.,
%
% we have $\var{g'}=\var{g}$ for every input $g'$ of a $\vee$-gate $g$.
%
%
%
%
%
%
%
%
%
Further, when solving WMC with weights defined in a ring,
we can simply impose determinism and decomposability without NNF: this leads to \emph{d-Ds}~\cite{monet2018towards}.

Determinism can also help for
enumeration tasks: 
on a d-DNNF $C$,  we can efficiently enumerate 
satisfying valuations 
in increasing order of 
a semiring weighting of the
literals~\cite{bourhis2022pseudo,amarilli2024ranked}, or uniformly sample 
satisfying valuations in $O(\mathrm{depth}(C) \times |\var{C}|)$ after a
preprocessing in $O(|C|)$~\cite{sharma2018knowledge}.

\paragraph*{Decision.}
One limitation of determinism is that it is a semantic property that may be hard to check on arbitrary circuits. Indeed, given a DNNF $C$ and a $\vee$-gate $g$ of $C$, it is \coNP-hard to check whether $g$ is deterministic.
This is not 
always a problem as circuits are sometimes deterministic by construction; 
but sometimes we prefer to enforce a sufficient condition for determinism, called \emph{decision gates}, which is easy to check syntactically.
A decision gate intuitively tests the value of one variable $x \in X$ before
proceeding further.
More precisely, a $\vee$-gate $g$ is a
\emph{decision gate} on~$x$ if the inputs of $g$ are exactly two $\wedge$-gates $g_0,g_1$, where
$g_0$ has one input labeled by $\neg x$ and $g_1$ has one input labeled by~$x$.
In other words, $g$ is of the form $(\neg x \wedge g_0) \vee (x \wedge g_1)$;
note that the literals $\neg x$ and~$x$ ensure that~$g$ is deterministic.
For instance, the leftmost $\vee$-gate of \Cref{fig:dnnf} is a decision gate on~$x_3$. %
A \emph{decision DNNF (dec-DNNF)} is a DNNF where every $\vee$-gate is a
decision gate. 
All dec-DNNFs are d-DNNFs, and their depth
can always be reduced to~$O(|X|)$ in linear time.
A circuit that only has decision gates is often called a \emph{binary decision diagram} or a \emph{branching program}: this includes in particular
\emph{Ordered Binary Decision Diagrams} (OBDDs)~\cite{bryant1986obdd}. %
\mbox{OBDDs} can be seen as DNNFs having only decision gates
and having an order $<$ on~$X$ such that decision gates testing a variable $x_i$ must come before gates testing variables $x_j$  whenever $x_i < x_j$.
OBDDs without this order requirement but with the analogue of decomposability are called \emph{Free Binary Decision Diagrams} (FBDDs).
See~\cite{amarilli2024circus} for a recent survey on the relationships between circuit classes and diagram classes.

\paragraph{Structuredness.} Structuredness~\cite{pipatsrisawat2008structure} is a generalization of orders in OBDD tailored for DNNFs. For a variable set $X$, a \emph{variable tree} or \emph{v-tree} for $X$ is a 
full binary tree whose leaves are in one-to-one correspondence with $X$.
Given a DNNF circuit $C$ on variables $X$ and a v-tree $T$ for~$X$, we say that a $\wedge$-gate~$g$ in $C$ \emph{respects $T$} if $g$ has exactly two inputs $g_1, g_2$ and if there is a node $t$ of $T$ with inputs $t_1,t_2$  such that $\var{g_1} \subseteq \var{t_1}$ and $\var{g_2} \subseteq \var{t_2}$ where $\var{t_i}$ is the set of variables in the leaves of the subtree of~$T$ rooted at $t_i$. A DNNF $C$ \emph{respects} $T$ if every $\wedge$-gate of $C$ does.
We call~$C$ \emph{structured} (denoted as SDNNF) if it respects some v-tree.

In essence, structuredness restricts how $\wedge$-gates are allowed to split the variables.
It naturally appears in many algorithms building DNNFs from other Boolean function representations, e.g., building a DNNF from a bounded-treewidth circuit~\cite{amarilli2020connecting, bova2015compiling}.
Structuredness unlocks two new results: first, 
there is an FPRAS to approximately count the satisfying valuations of
(non-deterministic) SDNNFs~\cite{arenas2021approximate}.
Second, 
enumerating the satisfying valuations of a d-SDNNF $C$ can be done with 
preprocessing $O(|C|)$ and output-linear delay $O(|X|)$ ~\cite{amarilli2017circuit}.

\begin{table}
  \caption{Tractable tasks for circuit classes. The input circuit $C$ is of size
  $s$, depth $d$, and $n$ variables; $k$ is the number of solutions to output
  (for Enum and Sampling).
}
    \label{fig:kc-task}
  \begin{tabu}{l@{~~~~}l@{~~~~}l@{\!\!}r}
    \toprule
    \rowfont[l]{\bf } Task & Circuit class & Complexity & \hfill Ref.\tabularnewline \midrule
    WMC & d-DNNF & $O(ns)$&~\cite{kimmig2017algebraic}  \tabularnewline
    WMC in ring & d-D & $O(ns)$&~\cite{monet2018towards}  \tabularnewline
    ApproxWMC & DNF & FPRAS&~\cite{karp1989monte,suciu2011probabilistic} \tabularnewline
    ApproxWMC & SDNNF & FPRAS&~\cite{arenas2021approximate} \tabularnewline
    Enum & d-SDNNF & $O(s+nk)$&~\cite{amarilli2017circuit}  \tabularnewline
    Sampling & d-DNNF & $O(s+dnk)$&~\cite{sharma2018knowledge} \tabularnewline
    \bottomrule
  \end{tabu}
\end{table}

\subsection{Tools}
\label{sec:tools}

An attractive feature of 
knowledge compilation is that 
many of its algorithms have implementations (at least experimental ones).
We now survey these practical tools.

There are two main families of knowledge compilation tools. 
First, 
\emph{top-down} tools are based on a generalization of the DPLL algorithm, known as exhaustive DPLL~\cite{bacchus2003algorithms}. It is based on a recursive procedure 
originally devised for solving \#SAT but which 
implicitly compiles into dec-DNNF formulas~\cite{huang2005dpll}. It compiles a CNF formula $\phi$ as follows: pick some variable $x$, recursively compile a circuit with  gates $g_0$ computing $\phi[x:=0]$ and $g_1$ computing $\phi[x:=1]$, and add a decision variable $g$ on $x$ connected to $g_0$ and $g_1$. When $\phi = \phi_1 \wedge \phi_2$ with $\phi_1, \phi_2$ on disjoint variables, the algorithm compiles them independently and connects the two circuits with a decomposable $\wedge$-gate. 
Efficient algorithms rely on two main ingredients: (1) a \emph{caching mechanism} that remembers previously-computed formulas to reuse them elsewhere in the circuit;
and (2) a \emph{heuristic} to choose variables to branch on, 
e.g., to break down the formula into smaller connected components. The knowledge compiler {\textsf d4}~\cite{d4source,lagniez2017d4} implements this algorithm, trying, e.g.,
to find balanced cutsets in the formula and 
performing oracle calls to SAT solvers to cut unsatisfiable branches.
The \#SAT solver \textsf{SharpSat-TD}~\cite{SharpSat-TDsource, korhonen2021sharpsattd} uses a heuristic that is guided by a tree decomposition computed via the FlowCutter algorithm~\cite{hamann2018flowcutter,strasser2017tdflowcutter}. It has recently been modified into a knowledge compiler~\cite{SharpSatKCsource, kiesel2023sharpsattd}. 

Second, \emph{bottom-up knowledge compilers}
take a CNF formula $\phi$ and
build circuits computing each clause of~$\phi$, before combining  them
into a bigger circuit computing $\phi$. This 
works if the target circuit class efficiently supports conjunction: combine two circuits $C_1,C_2$ into a circuit computing $C_1 \wedge C_2$.
Libraries such as
\textsf{CuDD}~\cite{somenzicudd} use this approach to compile CNFs to OBDDs.
Further, the knowledge compiler SDD~\cite{SDDsource, darwiche2011sdd} compiles into a subclass of structured d-DNNFs: these, unlike OBDDs~\cite{razgon2014obdds,bova2017compiling}, can efficiently handle
any bounded-treewidth CNF formula. 

%%% Local Variables:
%%% mode: latex
%%% TeX-master: "main"
%%% End:

\section{Circuits for MSO Queries over Trees}
\label{sec:mso}

We have seen in Section~\ref{sec:bool-funct-datab} how database tasks can be expressed
in terms of Boolean functions, and seen in Section~\ref{sec:kc} how such
functions can be represented as tractable circuits. We now start surveying how
circuit-based methods can be used to solve database tasks. We start in this
section by queries in \emph{monadic second-order logic} (MSO) over tree-shaped
data. This covers in particular the evaluation of word automata over
textual documents, including the so-called \emph{document
spanners}~\cite{fagin2015document}; and the evaluation of MSO queries over
bounded-treewidth data via Courcelle's theorem~\cite{courcelle1990monadic}.
We first give brief definitions of this setting, then study PQE for Boolean MSO queries
and enumeration tasks for MSO queries with free variables. Throughout this
section we adopt the \emph{data complexity} perspective, i.e., the MSO query is
always fixed, and the complexity is always a function of the input instance. 
We will study aggregation tasks (specifically, PQE) and enumeration tasks for
MSO queries over trees. In both settings, the results will proceed by
constructing tractable circuits to represent the Boolean provenance (for PQE) or the
answer function (for enumeration).

\paragraph*{Preliminaries.}
We consider queries over \emph{$\Sigma$-trees}, or simply \emph{trees}, which consist of nodes labeled with a
fixed alphabet $\Sigma$. We assume trees to be rooted, ordered, binary, and full. The
queries that we run over trees are expressed in MSO:
this language extends first-order logic with quantification over sets, on a
signature where we can test the label of tree nodes and the child and parent
relationship between tree nodes. For example, we can express in MSO that there
are two incomparable nodes with a certain label, or that nodes with a certain
label are totally ordered by the descendant relation. An MSO query~$Q$ may be
Boolean, in which case it can equivalently be expressed as a \emph{tree
automaton}; or it may have free variables. We can always assume without
loss of generality that each free variable~$X$ is second-order, because $Q$
can assert if necessary that $X$ must be a singleton.

\paragraph*{Aggregation tasks.}
We start with aggregation tasks for Boolean MSO queries~$Q$, specifically 
probabilistic query evaluation (PQE).
The PQE problem for~$Q$ asks for the probability that $Q$ is satisfied on an
input \emph{probabilistic tree}, as we will define shortly.
From there, by~\cite{amarilli2015provenance}, using
Courcelle's theorem~\cite{courcelle1990monadic}, these
results generalize to PQE over
tuple-independent databases of \emph{bounded treewidth}; and bounded
treewidth is in some sense the
most general condition that ensures
tractability~\cite{amarilli2016tractable,amarilli2022weighted}.

Formally, to define 
probabilistic trees, we distinguish a \emph{default label} $e \in \Sigma$;
a probabilistic tree then consists of a $\Sigma$-tree $T$ with a
function~$\pi$
giving a probability $\pi(n)$ to each tree node $n$ of~$T$. The semantics is
that $\mathcal{T} = (T, \pi)$ represents a probability distribution on possible worlds which
are $\Sigma$-trees with same skeleton as~$T$: 
each node~$n$ either keeps its label in~$T$ with probability~$\pi(n)$, or
``reverts'' to the default label~$e$ with probability $1-\pi(n)$, all these
choices being independent. PQE for~$Q$ on~$\mathcal{T}$
then asks for the total probability of the possible worlds
of~$\mathcal{T}$ that satisfy~$Q$.

Following earlier results on probabilistic XML
\cite{cohen2009running}, it is then known~\cite{amarilli2015provenance} that PQE
for~$Q$ can be computed in PTIME in the input~$\mathcal{T}$.
Specifically, we can define as in Section~\ref{sec:bool-funct-datab} the
\emph{provenance} of~$Q$ on~$T$ as the Boolean function defined on the nodes
of~$T$ that maps each valuation to~$0$ or~$1$ depending on whether the
corresponding possible world satisfies~$Q$. One can then show, representing~$Q$
as a tree automaton~$A$, that the provenance can be computed in time $O(|A|
\times |T|)$ as an SDNNF whose v-tree follows~$T$. There is
also a correspondence between properties of~$A$ and of~$C$,
e.g., if $A$ is unambiguous then $C$ is a
d-SDNNF~\cite{amarilli2024circus}. The tractability of WMC
for \mbox{d-DNNF} then implies that PQE for any fixed MSO query
on probabilistic trees can be solved in PTIME.

\paragraph*{Enumeration tasks.}
We now study MSO queries with free
second-order variables $Q(X_1, \ldots, X_k)$, again running over trees.
We first consider the task of enumerating query answers, again in data
complexity. It is known that, in
this setting, the answers to~$Q$ on an input tree~$T$ can be enumerated with
preprocessing in~$O(|T|)$, and with delay which is output-linear, i.e.,
which depends only on the size of each produced answer. This was shown first by
Bagan~\cite{bagan2006mso}, then by Kazana and Segoufin (only for free first-order
variables)~\cite{kazana2013enumeration}. This result can be recaptured via
knowledge compilation: a variant of provenance computation
allows us to obtain d-SDNNF representations of the answer functions
of MSO queries~\cite{amarilli2017circuit}. We can then enumerate the results of
the query with linear preprocessing and output-linear delay with an algorithm to
enumerate the satisfying valuations for this circuit
class~\cite{amarilli2017circuit}.

Similar results have also been shown for \emph{document spanners}, which essentially amounts to evaluating MSO
queries specified by automata over words, in particular via
algorithms that also ensure tractability in the input automaton~\cite{amarilli2019constant,amarilli2019enumeration}. In more
expressive settings, the work of Muñoz and
Riveros~\cite{munoz2022streaming,munoz2022constant} uses so-called
\emph{enumerable compact sets}, which resemble d-DNNF circuits, to achieve
efficient enumeration on \emph{nested documents} and \emph{SLP-compressed
documents}; and these are also used in
\cite{amarilli2022efficient} to enumerate the results of
\emph{annotation grammars}.

Enumeration algorithms for MSO over words and trees have been further extended
to \emph{ranked enumeration}: first by Bourhis et al.~\cite{bourhis2021ranked}
on words, with weights defined by MSO cost functions; then on
trees~\cite{amarilli2024ranked}, with weights defined on partial assignments by
so-called subset-monotone ranking functions. The latter work explicitly uses the
circuit approach via smooth multivalued d-DNNF circuits.

We last mention the \emph{incremental maintenance} of enumeration structures for
MSO queries on trees. The point of such structures is to enumerate the answers
of MSO queries while supporting updates to the underlying data; we want to
handle 
updates (and restart the enumeration) without re-running the
preprocessing phase from scratch. 
Note that this generalizes the task of incrementally maintaining Boolean MSO
queries~\cite{balmin2004incremental}.
In this setting, for \emph{relabeling} updates to the underlying tree, the best
known bounds are obtained via d-DNNF representations of answer
functions~\cite{amarilli2018enumeration}, improving on the work by Losemann and
Martens~\cite{losemann2014mso}. Specifically,
\cite{amarilli2018enumeration}
shows that enumeration with linear preprocessing and output-linear delay can be extended to
support relabeling updates in logarithmic time; tractability in the automaton
is also possible~\cite{amarilli2019constant}.

\section{Aggregative Tasks for CQs and UCQs}
\label{sec:dnnf-db}

Having shown the uses of circuits for MSO queries on trees, we now move 
to Boolean CQs and UCQs over arbitrary data.
We focus on aggregation tasks and study enumeration tasks in
Section~\ref{sec:query_answers}.
We first focus on
\emph{conjunctive queries} (CQs), which are existentially quantified
conjunctions of atoms; and on \emph{unions of conjunctive queries} (UCQs).
We show how circuits can be used for
\emph{probabilistic query evaluation} (PQE) on tuple-independent databases
(TIDs), and its special case
\emph{uniform reliability} (UR).
We first concentrate on \emph{exact} (i.e., non-approximate) PQE, and we study
it in data complexity, i.e., for fixed queries~$Q$: first for CQs under a \emph{self-join-freeness} assumption, then for
UCQs. We then move to approximate PQE, and to \emph{combined complexity} where
both~$Q$ and the TID are given as input. Last, we
cover Shapley values as another aggregation task.

\paragraph*{Exact PQE for self-join-free CQs.}
To study PQE, one first class of
queries to consider are the so-called \emph{self-join-free CQs}
(SJFCQs). A CQ is \emph{self-join-free} if each relation occurs only once,
i.e., there are no two atoms with the same symbol. For such queries, a
dichotomy on PQE was shown by Dalvi and Suciu~\cite{dalvi2007efficient}: a SJFCQ is either
\emph{hierarchical}, in which case PQE is in polynomial-time data complexity; or
it is \emph{non-hierarchical}, in which case PQE is \#P-hard in data complexity
(and so is UR~\cite{amarilli2022uniform}).
The tractability of hierarchical SJFCQs can be explained via tractable circuit
representations of the Boolean provenance, which is called the
\emph{intensional} approach to PQE. More specifically, for hierarchical SJFCQs, the
provenance can be computed in polynomial-time as an OBDD, in fact even as a
read-once formula~\cite{OlHu08}.

\paragraph*{Exact PQE for UCQs.}
Following this dichotomy on PQE for SJFCQs, Dalvi and Suciu have shown a far
more general dichotomy on UCQs: the PQE problem enjoys
PTIME data complexity for some UCQs (called \emph{safe}), and for all others
(the \emph{unsafe} UCQs) there is a \#P-hardness result for PQE~\cite{dalvi2012dichotomy},
indeed even for UR for most unsafe UCQs~\cite{kenig2021dichotomy}.
Ten years later, understanding this dichotomy in terms of tractable circuits is still
an open research problem. Indeed, the algorithm of~\cite{dalvi2012dichotomy}
follows the so-called \emph{extensional} approach for PQE and directly computes the
probability of the query. It does not follow the \emph{intensional
approach} of going via provenance. The \emph{intensional vs.\ extensional
conjecture}~\cite{monet2020solving} thus asks whether we can solve PQE for any
safe UCQ by
computing a provenance representation in a tractable circuit class and invoking
WMC on that class.

The intensional vs.\ extensional conjecture was investigated first
by Jha and Suciu~\cite{jha2012knowledge}: they characterize the strict subset of
safe UCQs whose provenance can be expressed as read-once formulas (like for
hierarchical SJFCQs), and also
the larger strict subset, called \emph{inversion-free UCQs}, for which we can
build OBDDs in PTIME. It was then shown in~\cite{bova2017circuit} that safe UCQs that
are not inversion-free do not admit polynomial-size provenance representations
even as \mbox{d-SDNNFs}. Jha and Suciu~\cite{jha2012knowledge}
also give sufficient conditions on safe UCQs to
admit polynomial-size \mbox{FBDD} provenance representations, but without
a characterization. It
was later shown in~\cite{beame2017exact} that some safe UCQs admit no
polynomial-size provenance representation as so-called DLDDs, implying the
same for dec-DNNFs and FBDDs.

However, the intensional vs.\ extensional conjecture is still open for more
expressive circuit classes with tractable WMC. It remains open whether the
class of d-DNNFs can tractably represent the provenance of all safe UCQs
(with~\cite{jha2012knowledge} conjecturing that it does not). 
The question is also open for the more
general class of d-Ds, which is in fact not yet separated from
d-DNNFs~\cite{monet2020solving}. The ability to compute polynomial d-Ds
to represent the provenance of all safe UCQs currently hinges on the unproven
\emph{non-cancelling intersections conjecture}~\cite{amarilli2024non}.

\paragraph*{Approximate PQE for UCQs and combined complexity.}
Faced with the general intractability of PQE for unsafe UCQs, it is natural to
settle for approximate PQE. Additive
approximations 
can be
obtained simply via Monte Carlo sampling~\cite{suciu2011probabilistic}, and
multiplicative approximations can always be obtained through the
intensional approach. Namely, for any fixed UCQ, we can represent its provenance 
as a monotone DNF in PTIME data complexity, and we can then solve
approximate PQE by solving approximate weighted model counting (ApproxWMC) on the DNF via
the Karp-Luby algorithm~\cite{karp1989monte} (see \Cref{sec:kc}), giving an FPRAS for the task.

This tractability of approximate PQE leads to the question of finding efficient
algorithms in combined complexity, i.e., when the query is also given as
input. In this light, 
Van Bremen and Meel~\cite{bremen2023pqe}
have studied the
combined complexity of SJFCQs of \emph{bounded hypertreewidth}, which are tractable for non-probabilistic query
evaluation~\cite{gottlob2002hypertree},
and extended this tractability result
to approximate PQE. 
We will explain in \Cref{sec:query_answers} how their algorithm
can be understood via SDNNF
circuits and via the FPRAS of~\cite{arenas2023complexity} mentioned in \Cref{sec:kc}.
Note that \cite{bremen2023pqe} generally does not extend to CQs with
self-joins~\cite{amarilli2024conjunctive}.
We also note that provenance-based
approaches have also been used 
for tractable combined algorithms for \emph{exact PQE}, e.g., via $\beta$-acyclic positive
DNFs~\cite{amarilli2017conjunctive,brault2015understanding}.

\paragraph*{Shapley values.}
We can use circuits for other aggregation tasks than PQE: one important
example is computing the \emph{Shapley value} of facts in
relational instances. This task has been shown to be
tractable whenever we can compute a
representation of the query provenance as a
d-D~\cite{deutch2022computing}, and the same was shown for the
related notion of Banzhaf values~\cite{abramovitch2023banzhaf}.
As another example of an aggregation task, the computation of the Shapley
value (and variants)
has also been recently extended to the setting of probabilistic
instances~\cite{karmakar2024expected}, also using d-D circuits.

\paragraph*{Implementation.}
We last mention 
\emph{ProvSQL}~\cite{senellart2018provsql}, as it is a concrete instantiation of
the circuit-based approach to provenance~\cite{deutch2014circuits}. ProvSQL is a module of the
\mbox{PostgreSQL} relational database management system, which adds the possibility to
track the provenance of query results as a circuit throughout query evaluation.
These general-purpose circuits can then be used for various applications. In
particular, they can be used for PQE 
via existing knowledge compilation tools for WMC (see \Cref{sec:tools}), or for
Shapley value
computation~\cite{karmakar2024expected}.

\section{Circuits to Represent Query Answers}
\label{sec:query_answers}

We now move to a different perspective on circuits, which we
will use in particular for enumeration tasks with CQs.
We do not use answer functions for CQs, because such queries cannot easily
express that a variable is assigned to \emph{exactly one value}.
For this reason, unlike previous sections, we will go beyond Boolean functions
and Boolean circuits, and adopt a \emph{multivalued} perspective: 
we will show how circuits can %
succinctly represent \emph{relations}, in particular
query answers.

\paragraph{Representing Relations as Circuits.}

Let $f \from \{0,1\}^X \rightarrow \{0,1\}$ be a Boolean function. We can easily
see $f^{-1}(1) \subseteq \{0,1\}^X$ as a relational table~$R$, whose domain is
$\{0,1\}$ and whose attributes (in the named perspective) is~$X$; see, e.g., the table right of \Cref{fig:dnnf}. 
What is more, we can see circuit representations of~$f$ as a factorized
representation of~$R$.
More specifically, a \mbox{DNNF}~$C$ on variables $X$ can be seen as a factorized way of building up a relation from elementary relations of the form $x=0$ or $x=1$. 
Then, $\wedge$-gates are a natural join of their input relations (denoted
by~$\bowtie$), while $\vee$-gates are unions.

Unfortunately, 
the semantics of $\vee$-gates does not precisely correspond to the union
operator of relational algebra.
Indeed, the union operator applies to tables with the same attributes, whereas two Boolean
functions $f$ and $g$ on different variable sets $X \neq Y$ can be disjoined as
$f\vee g$, giving a function on variables $X \cup Y$. 
This issue does not arise with \emph{smooth} circuits 
(see \Cref{sec:kc}), but to interpret $\vee$ for general circuits we must
extend the union operator.
Formally, given two relations $R \subseteq D^X$ and $S \subseteq D^Y$, the
\emph{extended union} $R \ecup S$ of $R$ and $S$ is the relation on attributes
$X \cup Y$ defined as $(R \times D^{Y \setminus X}) \cup (S \times D^{X \setminus Y})$\footnote{Note that, unlike union,
the extended union depends on the domain~$D$. An alternative choice is to fix a
default value $d$ and define the union as 
$R \ecup' S = (R \times \{d\}^{Y \setminus X}) \cup (S \times \{d\}^{X \setminus
Y})$. When $D = \{0,1\}$, the choice of
$d=0$ corresponds to the \emph{zero-suppressed semantics} studied for
decision diagrams~\cite{minato1993zero,wegener2000branching}.}. If $X=Y$ then we clearly have $R \ecup S = R \cup S$. 

We can now directly generalize NNFs and their variations to \emph{relational
circuits} on any finite domain: see \Cref{fig:boolean-vs-relational} for a summary.
Namely, a $\{\ecup, \bowtie\}$-circuit $C$ on attributes $X$ and domain $D$ is a circuit whose internal gates are labeled by either $\ecup$ or $\bowtie$ and whose input gates are labeled by relations of the form $x/d$ where $x \in X$ and $d \in D$.
The \emph{attributes} $\attr{g} \subseteq X$ of a gate $g$ of $C$ are the attributes 
labeling the inputs of $C$  that have a directed path to~$g$. Further, $g$ computes a relation $\rel{g} \subseteq D^{\attr g}$ inductively defined from the relations computed by its inputs and from the label of~$g$.
The circuit~$C$ computes $\rel{C}$ which is $\rel{o}$ for~$o$ the output of~$C$, again extended by allowing arbitrary values in~$D$ for the attributes of $X \setminus \attr{o}$.
One can easily check that if $C$ is a $\{\ecup,\bowtie\}$-circuit on domain $\{0,1\}$, then we can get an NNF computing $\rel{C}$ by renaming every input of the form $x/1$ by $x$ and $x/0$ by $\neg x$ and replacing every $\bowtie$-gate by a $\wedge$-gate and every $\ecup$-gate by a $\vee$-gate.

Now, if a $\bowtie$-gate has inputs whose attributes are pairwise disjoint, it 
actually computes the Cartesian product of these inputs, and we can denote it by $\times$
and call it \emph{decomposable}.
Hence, a $\{\ecup, \times\}$-circuit is a $\{\ecup, \bowtie\}$-circuit where
every $\bowtie$-gate is in fact computing a Cartesian product. Such
\emph{decomposable} circuits 
correspond to DNNFs when restricted to the Boolean domain. Similarly, if a $\ecup$-gate $g$ has inputs $g_1,\dots,g_k$ such that $\attr{g_1} = \dots = \attr{g_k}$, it actually computes a union and we denote it by $\cup$. This gives $\{\cup, \times\}$-circuits, which correspond to smooth DNNFs. 
We can also generalize structuredness:
a $\times$-gate $g$ with two inputs $g_1,g_2$ \emph{respects} a vtree $T$ on~$X$ if there is a node $t$ in $T$ such that $\attr{g_1} \subseteq \attr{t_1}$ and $\attr{g_2} \subseteq \attr{t_2}$ where $t_1,t_2$ are the children of $t$ in $T$ and each $\attr{t_i}$ is the set of attributes labeling the leaves of the subtree of $T$ rooted at~$t_i$.

We can also generalize determinism: denoting disjoint union by~$\uplus$,
we have $\{\uplus,\times\}$-circuits which correspond to smooth d-DNNFs.
Again, the disjointness of $\cup$-gates is 
a semantic property that may be intractable to verify.
Like in \Cref{sec:kc}, we can enforce it by a sufficient syntactic condition:
a $\uplus$-gate is a \emph{decision gate} if it is of the form $\biguplus_{d \in
D'} [x/d] \times g_d$ for a subset $D' \subseteq D$, where 
the $[x/d]$ are input gates.
A \emph{$\{\dec,\times\}$-circuit} is a 
$\{\cup,\times\}$-circuit where every $\cup$-gate is a decision gate.

\begin{table}
\caption{From Boolean circuits to relational circuits.}
  \label{fig:boolean-vs-relational}
  \begin{tabu}{X@{~~}c@{~~}l}
    \toprule
    \rowfont[l]{\bf } Boolean Circuits &
    \multicolumn{2}{l}{~~~~~~~~~Relational
    Circuits} \tabularnewline \midrule
    NNF & $\{\ecup, \bowtie\}$ & \tabularnewline
    DNNF & $\{\ecup, \times\}$ & or $\{\cup, \times\}$ if smooth \tabularnewline
    d-DNNF & $\{\euplus, \times\}$ & or $\{\uplus, \times\}$ if smooth  \tabularnewline
    dec-DNNF & $\{\dec,\times\}$ &  \tabularnewline
    \bottomrule
  \end{tabu}
\end{table}

\paragraph{Factorized Databases.}
The notion of relational circuits that we present here is related to 
\emph{factorized databases} as introduced by Olteanu and Z{\'a}vodn{\`y}~\cite{olteanu2012factorised, olteanu2015size}.
Their \emph{$d$-representations} correspond to
$\{\cup,\times\}$-circuits, and their \emph{$f$-representations} further require
that the circuit 
is actually a tree, i.e.,
there is no sharing.
Further, many 
factorized database algorithms follow a tree over the database attributes, so
that they build structured $\{\dec, \times\}$-circuits. The link between factorized databases and 
knowledge compilation 
is also explored in~\cite{olteanu2016factorized}.

\paragraph{Building and Using Circuits.}

We now turn to the use of relational circuits to solve enumeration tasks for CQs.
Deciding whether a CQ has at least one answer is \NP-hard in
combined complexity~\cite{chandra1977optimal} and the same reduction shows that
counting query answers is \sP-hard. 
Thus, a fruitful line of research studies how to design algorithms that use the query
structure to enumerate or
count CQ answers efficiently.
For example, Yannakakis observed in~\cite{yannakakis1981algorithms} that one can
test with linear data complexity whether an \emph{acyclic} CQ has at least one answer: further,
his algorithm is also tractable in the query size.
This result has been generalized in many ways: to constant-delay
enumeration for the so-called \emph{free-connex acyclic CQs} by Bagan, Durand and Grandjean~\cite{bagan2007acyclic}; to counting
for projection-free CQs in~\cite{pichler2013tractable} and for
arbitrary CQs in~\cite{durand2014complexity};
 to direct access in~\cite{carmeli2022answering,bringmann2022tight} for some
 lexicographical orders; or to semiring-based aggregation queries in~\cite{abo2016faq,joglekar2016ajar}. %
Relational circuits offer a unifying view of all these results and more, thanks to
a claim initially shown in the setting of factorized databases:

\begin{theorem}[\!\cite{olteanu2015size}]\label{thm:CQ:compile}\!\!Given a free-connex acyclic conjunctive query $Q$ and a database $\db$, we can build a structured $\{\dec,\times\}$-circuit in time $O(\poly |Q|\cdot|\db|)$ computing $Q(\db)$. The circuit is of size $O(\poly |Q|\cdot|\db|)$.
\end{theorem}

This result is actually also shown for non-acyclic CQs
in~\cite{olteanu2015size}. In this case, the transformation will not be linear
wrt $|\db|$ anymore but of the form $O(|\db|^k)$ for $k$ a constant that depends
only on the structure of $Q$. When $Q$ is in the class of free-connex acyclic
CQs,
we have $k=1$. Otherwise, $k$ intuitively measures how far $Q$ is from being free-connex acyclic~\cite{fischl2018general};
if $Q$ is projection-free it is at most, e.g., the hypertreewidth of~$Q$~\cite{gottlob2002hypertree}.

There are two main techniques to  prove Theorem~\ref{thm:CQ:compile}, that is, to build circuits of linear size for free-connex acyclic conjunctive queries. The first one can be seen as
the trace of Yannakakis's algorithm~\cite{yannakakis1981algorithms} and is
intuitively the approach taken in~\cite{olteanu2015size}. This result decides
whether $Q(\db)$ is empty by alternatively joining and projecting the atoms of
$Q$ in an order that can be obtained from the fact that $Q$ is free-connex acyclic (see~\cite[Chapter
6]{libkin2004elements}).
This ensures that all intermediate results have size $O(|\db|)$. The trace of the
joins and projections done during the evaluation corresponds to a $\{\dec,\times\}$-circuit 
of size $O(\poly |Q| \cdot |\db|)$
computing $Q(\db)$. 

Another approach to build a circuit for $Q(\db)$ is inspired by the DPLL algorithm from top-down knowledge compilers~\cite{bacchus2003algorithms}. The circuit is built by following a static order on the attributes and recursively compiling $Q[x_1=d]$ for every value $d \in D$, which corresponds to building a decision gate on $x_1$. If the query can be written as $Q_1 \wedge Q_2$ at some recursive call with $\attr{Q_1} \cap \attr{Q_2} = \emptyset$, then $Q_1$ and $Q_2$ are compiled independently and their subcircuits are joined by a $\times$-gate. 
By choosing an order that witnesses the free-connex acyclicity of $Q$, and adding a caching mechanism to remember previously computed queries, one can show that this algorithm 
also
builds a $\{\dec,\times\}$-circuit of size $O(\poly |Q| \cdot |\db|)$ computing $Q(\db)$. If another order is chosen, the output of the algorithm is still a circuit but its size may not be linear in $|\db|$.
The detailed algorithm is given in~\cite{capelli2024directaccess}
with order-dependent complexity bounds ; it is
also exemplified in~\cite{dpllex}. 

The tractable tasks from \Cref{fig:kc-task} can straightforwardly be generalized to $\{\dec,\times\}$-circuits.
In particular, if $n$ is the number of attributes of $C$, $|\rel C|$ can be computed in time $O(n|C|)$ and $\rel C$ can be enumerated with delay $O(n)$: this fact was observed in~\cite{olteanu2015size} and used to motivate factorized databases. In particular, in data complexity, the value $n$ is a constant, hence we recover the result from~\cite{bagan2007acyclic} that $Q(\db)$ can be enumerated with constant-delay data complexity (more precisely $O(\poly |Q|)$) after a $O(\poly |Q| \cdot |\db|)$ preprocessing when $Q$ is free-connex acyclic, the preprocessing being here the construction of the circuit from \Cref{thm:CQ:compile}.
We note that, conditionally, for self-join-free conjunctive queries, it is known that only
free-connex acyclic queries enjoy this tractability
guarantee~\cite{bagan2007acyclic}. This conditional lower bound does not extend
to CQs with self-joins~\cite{carmeli2023conjunctive}, nor to UCQs~\cite{bringmann2022tight}.

More interestingly, the DPLL-based construction naturally produces $\{\dec,\times\}$-circuits of a very particular form: the circuit is built in a way where there exists an order $\{x_1,\dots,x_n\}$ on the attributes of $Q$ such that, for every decision gate $g$ on $x_i$ with input $g_1,\dots, g_k$, we have $\attr{g_p} \subseteq \{x_{i+1}, \dots, x_n\}$. This makes it possible, after a preprocessing time of $O(|C|)$, to do direct access for the lexicographical order induced by $x_1,\dots,x_n$ in time $O(n \cdot \log |\db|)$~\cite{capelli2024directaccess}. Combined with \Cref{thm:CQ:compile}, it is an alternative proof of the results of~\cite{carmeli2022answering}. The approach also generalizes to non-acyclic queries, matching results from~\cite{bringmann2022tight}, and to 
so-called \emph{signed conjunctive queries}.

\paragraph{From Relational Circuits to Provenance.}

Representing answers of CQs with relational circuits is close to computing provenance, and has actually been one of the original motivations~\cite{zavodny2011factorisation}. We make this intuition formal by 
explaining how relational circuits can be used to compute the Boolean provenance of queries.

Consider a signature $\sigma$ and a
projection-free CQ $Q$ (that is, every variable of $Q$ is free).
We first define a signature $\sigma^*$: for each relation $R$ in $\sigma$ we add
a relation $R^*$ in~$\sigma^*$ with arity increased by~1. We then denote by
$Q^*$ a query on~$\sigma^*$ obtained as follows: for each atom $A = R(\tup)$ of
$Q$, add an atom $R^*(\tup,y_A)$ to~$Q'$, where $y_A$ is a fresh attribute for the atom~$A$.
Given an instance $\db$ on~$\sigma$,
we also let $\db^*$ be the instance on~$\sigma^*$
where each symbol $R^*$ is interpreted as $R$ in $\db$ but each fact~$F$ is augmented with a fresh value $a_F$.
It is easy to check that $Q(\db)$ and $Q^*(\db^*)$ are isomorphic up to the added identifiers. Moreover, if $Q$ is acyclic, then $Q^*$ is also acyclic
(they even have the same hypertreewidth).

Now, let $C$ be a $\{\cup,\times\}$-circuit computing $Q^*(\db^*)$.
Its inputs are of two kinds: relations of the form $x/a$ for $x$ an attribute of~$Q$
and $a$ a value of~$\db$, and relations of the form $y_A/a_F$ where $a_F$ is a fresh value 
added for the tuple~$F$.
We modify~$C$ in two steps. First, let~$C'$ be the circuit obtained from~$C$ by
replacing each input $y_A/a_F$ of the second kind by a 
variable $X_F$ on domain $\{0,1\}$.
Note that $C'$ is generally not decomposable, but decomposability is preserved
if $Q$ is self-join-free:
for each fact $F$ of~$\db$, the variable $X_F$ of~$C'$ then corresponds to only
one input label of~$C$, namely inputs of the form $y_A/a_F$ 
where $A$ is the one atom of~$Q$ for the relation used by~$F$.
In any case, a Boolean valuation $\nu$ of the variables $\{X_F \mid F \in \db\}$ can be naturally identified as a subinstance $\db_\nu \subseteq \db$ 
containing the facts $F$ of~$\db$ such that $\nu(X_F) = 1$, and
we can then easily see that $C'$ computes $\{ \tau \times \nu \mid \nu \in 2^\db, \tau \in Q(\db_\nu)\}$. 
Second, let $C''$ be the circuit obtained from~$C'$ by existentially projecting
away every variable of the first kind, i.e., the variables that are not of the
form~$X_F$.
If $C'$ is decomposable then $C''$ also is; however $C''$ is generally not
deterministic.
Hence, $C''$ is a $\{\cup,\times\}$-circuit computing $\{ \nu \mid \nu \in 2^\db \text{ s.t. }
 Q(\db_\nu) \neq \emptyset \}$.
That is, it computes the Boolean provenance of the Boolean query obtained by
existentially quantifying~$Q$. Further, $C''$ has size less than $|C|$. Last, if $Q$ is self-join-free then $C''$ is a DNNF, and $C''$ is structured if~$C$ was.

Among other things, this result allows us to recover the result of Van Bremen
and Meel~\cite{bremen2023pqe} mentioned in \Cref{sec:dnnf-db}. Recall that
they give 
a combined FPRAS for
approximate PQE with self-join-free CQs~$Q$ 
of bounded hypertreewidth.
This is straightforward in the data complexity sense (see \Cref{sec:dnnf-db}), so the interesting point is the tractability in combined complexity.
Applying \Cref{thm:CQ:compile} 
(extended for queries of hypertreewidth $k$)
to the query~$Q^*$ defined as above, we get
a structured $\{\dec,\times\}$-circuit that computes $Q^*(\db^*)$ and which has size $O(\poly |Q| \cdot |\db|^k)$. As~$Q$ is self-join-free, our transformation above
gives 
a SDNNF $C$ computing the Boolean provenance of $Q$ of size $O(\poly |Q| \cdot |\db|^k)$. We can then use the FPRAS from~\cite{arenas2021approximate} on $C$ to solve approximate PQE for~$Q$ on~$\db$, which concludes.
%

%%% Local Variables:
%%% mode: latex
%%% TeX-master: "main"
%%% End:

\section{Perspectives}
\label{sec:perspective}

We have seen how results from database theory can often be
obtained via tractable circuit classes, or how some existing proofs can 
be rephrased in this vocabulary. We have focused on two main kinds of tasks:
\emph{aggregation tasks}, including PQE and
Shapley value computation; and \emph{enumeration tasks}, including ranked
enumeration and direct access. We have focused on the settings of MSO queries and
of CQs and UCQs. We believe that this survey illustrates the versatility of
circuit techniques, which can sometimes offer a unified view on the
tractability of multiple tasks in different areas. Circuits also
serve as a convenient intermediate language between database algorithms
(which deal with the query and data) and task-specific
algorithms (e.g., satisfiability, counting, etc.).
This view makes it possible to give modular algorithms and to leverage existing solver
implementations.

We have focused on circuit classes from knowledge compilation, but we
point out that circuits have been used in other interesting ways in database theory, for example, arithmetic circuits for semiring provenance~\cite{deutch2014circuits}, circuits for secure aggregation on data shared by different parties~\cite{wang2021secure}, or circuits for efficient parallel evaluation of queries~\cite{wang2022query,keppeler2023work}. Similarly, we have chosen to focus on query evaluation applications in databases, but tractable circuits are also used in closely related topics such as CSPs~\cite{carbonnel2020point, berkholz2024characterization}, homomorphism representations~\cite{berkholz2023dichotomy}, or the computation of SHAP-scores~\cite{arenas2023complexity}.

We believe that tractable circuits still have a lot to bring to database theory, and vice-versa; we conclude by highlighting some directions for future work. We see two main research axes: studying how far we can bring circuit methods in established contexts, and introducing them in new contexts. For the first axis, the \emph{intensional-extensional conjecture} for PQE of UCQs is a clearly identified setting where we do not know how far circuits can be taken to recapture existing results~\cite{monet2020solving,amarilli2024non}. A similar research direction would be the \emph{incremental maintenance of enumeration structures} for MSO on trees: while circuit methods handle substitution updates in logarithmic time~\cite{amarilli2018enumeration}, better complexities are possible, at least in the case of Boolean queries on words~\cite{amarilli2021dynamic}: it is unclear for now whether such results can be explained in terms of circuits.
It is also unclear which incremental results can be recaptured by circuits in the setting of incremental PQE~\cite{berkholz2021probabilistic}, incremental enumeration for CQs~\cite{berkholz2019constant,kara2020trade,kara2022conjunctive} or CQs with aggregates~\cite{idris2017dynamic,kara2020maintaining},
or more general results in algorithms on dynamic
data~\cite{hanauer2022recent}. Moreover, there remain tractability results for aggregate tasks in the database and CSP literature that have not been directly explained from a circuit perspective but which use similar counting techniques,
for example results on counting the number of answers of UCQs~\cite{focke2023counting,chen2016counting} or complex aggregation over semiring-annotated data~\cite{abo2016faq, joglekar2016ajar}.

For the second axis, we believe that circuits could be applied to entirely new areas.
One possibility is query enumeration for First-Order logic
(FO), e.g., over bounded-degree
structures~\cite{durand2007first,kazana2011first}. It is not known if such
results, and their subsequent
extensions~\cite{segoufin2017constant,schweikardt2022enumeration}, can be captured in terms of circuits.
A second setting in which circuits could be relevant is 
\emph{database repairs}, in particular counting
subset
repairs~\cite{kimelfeld2020counting,calautti2019counting,calautti2022counting},
enumerating them~\cite{kimelfeld2020counting}, and more generally
representing them in a factorized way.
One last question is to connect circuits to the study of efficient algorithms for
\emph{Datalog evaluation} including
provenance computation in various
semirings~\cite{abokhamis2024convergence}. Can tractable algorithms for these tasks
be connected to algorithms producing
tractable circuit representations? Can fine-grained complexity lower bounds be connected 
to circuit lower bounds?

\bigskip

{  {\bf Acknowledgements.} This work was supported by project ANR KCODA,
ANR-20-CE48-0004, and by project ANR CQFD, ANR-18-CE23-0003-02. We are
grateful to Tim van Bremen, to Benny Kimelfeld, and to Mikaël Monet for their insightful feedback.}

\newpage 
\balance
{
\small
\bibliographystyle{abbrv}
\bibliography{main}
}
\end{document}